\documentclass[copyright,creativecommons,noderivs,noncommercial]{eptcs}
\providecommand{\event}{WRS 2009} 
\providecommand{\volume}{??}
\providecommand{\anno}{2009}
\providecommand{\firstpage}{1}
\providecommand{\eid}{??}

 \usepackage{amsmath,amsfonts,amssymb}
 \usepackage{listings}
 \usepackage{verbatim}
 \usepackage{alltt}

\newtheorem{defn}{Definition}
\newtheorem{exmp}{Example}

\def\var\mathit{var}
\def\pif{\text{\tt :-}}
\def\rholog{$\rho$Log}
\def\prholog{P$\rho$Log}

\def\modein{\mathit{in}}
\def\modeout{\mathit{out}}

\def\cVa{\mathcal{V}_\mathsf{an}}

\def\cF{\mathcal{F}}

\def\vars{\mathit{vars}}

\def\map1{\mathit{map1}}

\def\map{\mathit{map}}

\title{Strategies in \prholog}
\author{Besik Dundua
\institute{RISC, JKU Linz, Austria}
\email{bdundua@risc.uni-linz.ac.at}
\and
Temur Kutsia
\institute{RISC, JKU Linz, Austria}
\email{kutsia@risc.uni-linz.ac.at}
\and
Mircea Marin
\institute{University of Tsukuba, Japan}
\email{mmarin@cs.tsukuba.ac.jp}
}
\def\titlerunning{Strategies in \prholog}
\def\authorrunning{B. Dundua, T. Kutsia, M. Marin}
\begin{document}
\maketitle
\lstset{language=Prolog}
\begin{abstract}
\prholog{} is an experimental extension of logic programming with strategic conditional transformation rules, combining Prolog with \rholog{} calculus. The rules perform nondeterministic transformations on hedges. Queries may have several results that can be explored on backtracking. Strategies provide a control on rule applications in a declarative way. With strategy combinators, the user can construct more complex strategies from simpler ones. Matching with four different kinds of variables provides a flexible mechanism of selecting (sub)terms during execution. We give an overview on programming with strategies in \prholog{} and demonstrate how rewriting strategies can be expressed.
\end{abstract}

\section{Introduction}
\prholog{} (pronounced P\={e}-r\={o}-log) is an experimental tool that extends logic programming with strategic conditional transformation rules, combining Prolog with \rholog{} calculus~\cite{MK06}. \rholog{} deals with hedges (sequences of terms), transforming them by conditional rules. Transformations are nondeterministic and may yield several results. Logic programming seems to be a suitable framework for such nondeterministic computations. Strategies provide a control on rule applications in a declarative way. Strategy combinators help the user to construct more complex strategies from simpler ones. Rules apply matching to the whole input hedge (or, if it is a single term, apply at the top position). Four different types of variables give the user flexible control on selecting subhedges in hedges (via individual and sequence variables) or subterms/contexts in terms (via function and context variables). As a result, the obtained code is usually quite short, declaratively clear, and reusable.

\prholog{} programs consist of clauses. The clauses either define user-constructed strategies by (conditional) transformation rules or are ordinary Prolog clauses. Prolog code can be used freely within \prholog{} programs. One can include its predicates in \prholog{} rules, which is especially convenient when arithmetic calculations or input-output features are needed.

\prholog{} inference mechanism is essentially the same as SLDNF-resolution, multiple results are generated via backtracking, its semantics is compatible with semantics of normal logic programs~\cite{Lloyd87} and, hence, Prolog was a natural choice to base \prholog{} on: The inference mechanism comes for free, as well as the built-in arithmetic and many other useful features of the Prolog language. Following Prolog, \prholog{} is also untyped, but values of sequence and context variables can be constrained by regular hedge or tree languages. We do not elaborate on this feature here.

For the users familiar with logic programming and Prolog it is pretty easy to get acquainted with \prholog{} and to quickly start writing programs, since its syntax is very similar to that of Prolog and semantics is based on logic programming.

We tried to provide as little as possible hard-wired features in the system to give the user a freedom in experimenting with different choices. Probably the most notable such feature is the leftmost-outermost term traversal strategy the \prholog{}'s matching algorithm uses, but it can also be easily modified since the corresponding Prolog code is open: Exchanging the order of clauses there would suffice. The user can also program different traversal strategies pretty easily inside \prholog{}.

The goal of this paper is to give an overview of \prholog{} and, in particular, show how it uses strategies. After briefly reviewing the related work in Section~\ref{sect:related}, we discuss the syntax of \prholog{} (Section~\ref{sect:prelim}) and then list some of the strategies from the library together with examples that explain the input-output behavior of the system (Section~\ref{sect:prog}). Next, we show how user-defined strategies can be introduced illustrating it on the examples of defining rewriting strategies in \prholog{} (Section~\ref{sect:rewriting}). One can see that the code there is quite short and readable, and it also demonstrates expressiveness of \prholog{}.

\section{Related Work}
\label{sect:related}

Programming with rules has been experiencing a period of growing interest since the nineties when rewriting logic~\cite{MartiOlietMesseguer2001} and rewriting calculus~\cite{CirsteaKirchner01a,CirsteaKirchner01b} have been developed and several systems and languages (ASF-SDF~\cite{vandenBrand:2001}, CHR~\cite{CHR}, Claire~\cite{Claire}, ELAN~\cite{ElanENTCS}, Maude~\cite{Maude}, Stratego~\cite{Visser01}, just to name a few) emerged. The \rholog{} calculus has been influenced by the $\rho$-calculus~\cite{CirsteaKirchner01a,CirsteaKirchner01b} as also its name suggests, but there are some significant differences: \rholog{} adopts logic programming semantics (clauses are first class concepts, rules/strategies are expressed as clauses), uses top-position matching, and employs four different kinds of variables. Consequently, \prholog{} (based on \rholog) differs from ELAN (based on $\rho$-calculus). Also, ELAN is a mature system with a very efficient compiler while \prholog{} is an experimental extension of Prolog implemented in Prolog itself.

From the architecture point of view, \prholog{} is closer to another mature system, CHR, because both extend the host language (in this case, Prolog) in a declarative way. However, the purpose of CHR is different: It has been designed as a language for writing constraint solvers. CHR extends Prolog with the rules to handle constraints that are the first class concept there. \prholog{} is not designed specifically for programming constraint manipulation rules and we have not experimented with specifying such rules.

In the OBJ family of languages (OBJ2~\cite{FutatsugiGJM85}, OBJ3~\cite{Goguen00OBJ}, CafeOBJ~\cite{FutatsugiN97}), local strategies can be explicitly specified. They guide evaluation: In function calls only the arguments, specified by the strategies are evaluated. Among the other systems, strategic programming is supported Maude, and Stratego. Maude is based on rewriting logic, can perform efficient pattern matching modulo equational theories like associativity, commutativity, idempotence, and has a high-performance rewrite engine. Stratego is a domain-specific language designed for program transformation, combining term traversal strategies, pattern matching, and rewriting.

To compare with these systems, \prholog{} has been designed with the purpose to experiment with strategic conditional transformation rules in a logic programming environment. Strategies and nondeterministic computations fit well into the logic programming paradigm. Bringing hedge and context pattern matching and strategic transformations into logic programs seems to facilitate writing declaratively clear, short, and reusable code.

\section{Preliminaries}
\label{sect:prelim}

\prholog{} is essentially based on the language of \rholog~\cite{MK06}, extending Prolog with it. Here we use the \prholog{} notation for this language, writing its constructs in \texttt{typewriter} font. The expressions are built over the set of functions symbols $\cF$ and the sets of individual, sequence, function, and context variables. These sets are disjoint. \prholog\ uses the following conventions for the variables names: Individual variables start with \verb#i_# (like, e.g., \verb#i_Var# for a named variable or \verb#i_# for the anonymous variable), sequence variables start with \verb#s_#, function variables start with \verb#f_#, and context variables start with \verb#c_#. The symbols in $\cF$, except the special constant \texttt{hole}, have flexible arity. To denote the symbols in $\cF$, \prholog{} basically follows the Prolog conventions for naming functors, operators, and numbers.

\emph{Terms} \texttt{t} and \emph{hedges} \texttt{h} are constructed in a standard way by the following grammars:
\begin{alltt}
     t ::= i_X | hole | f(h) | f_X(h) | c_X(t)
     h ::= t | s_X | eps | h_1, h_2
\end{alltt}
where \verb#eps# stands for the empty hedge and is omitted whenever it appears as a subhedge of another hedge. \texttt{a(eps)} and \verb#f_X(eps)# are often abbreviated as \texttt{a} and \verb#f_X#. A \emph{Context} is a term with a single occurrence of  \texttt{hole}. A context can be applied to a term, replacing the hole by that term. For instance, applying the context \verb#f(hole, b)# to \verb#g(a)# gives \verb#f(g(a), b)#.

A \emph{Substitution} is a mapping that maps individual variables to (hole-free) terms, sequence variables to (hole-free) hedges, function variables to function symbols, and context variables to contexts so that
\begin{itemize}
  \item all but finitely many individual, sequence, and function variables are mapped to themselves, and
  \item all but finitely many context variables are mapped to themselves applied to the hole.
\end{itemize}

The mapping is extended to arbitrary terms and hedges in the usual way. For instance, the image of the hedge \verb#(c_Ctx(i_Term), f_Funct(s_Terms1, a, s_Terms2))# under the substitution \verb#{c_Ctx#$\mapsto$\verb#f(hole), i_Term#$\mapsto$\verb#g(s_X), f_Funct#$\mapsto$\verb#g, s_Terms1#$\mapsto$\verb#eps, s_Terms2#$\mapsto$\verb#(b, c)} # is the hedge \verb#(f(g(s_X)), g(a, b, c))#.

In \cite{KM05}, an algorithm to solve \emph{matching equations} in the language just described has been introduced. Matching equations are equations between two hedges, one of which does not contain variables. Such matching equations may have zero, one, or more (finitely many) solutions, called matching substitutions or \emph{matchers}.

\begin{exmp}
The term \verb#c_X(f(s_Y))# matches \verb#g(f(a, b), h(f(a), f))# in three different ways with the matchers
 \verb#{c_X#$\mapsto$\verb#g(hole, h(f(a), f)), s_Y#$\mapsto$\verb#(a, b)}#,
 \verb#{c_X#$\mapsto$\verb#g(f(a, b), h(hole, f)),# \verb#s_Y#$\mapsto$\verb#a}#, and
 \verb#{c_X#$\mapsto$\verb#g(f(a, b), h(f(a), hole)), s_Y#$\mapsto$\verb#eps}#.

 The hedge \verb#(s_X, f_F(i_X, a, s_), s_Y)# matches \verb#(a, f(b), g(a, b), h(b, a))# with the matcher \verb#{s_X#$\mapsto$\verb#(a, f(b), g(a, b)), f_F#$\mapsto$\verb#h, i_X#$\mapsto$\verb#b, s_Y#$\mapsto$\verb#eps}#,
\end{exmp}

A \rholog\ \emph{atom} ($\rho$-atom) is a triple consisting of a term $\tt st$ (a \emph{strategy}) and two hedges $\tt h1$ and $\tt h2$, written as \verb#st :: h1 ==> h2#. (The hedges $\tt h1$ and $\tt h2$ do not contain the \verb#hole# constant.) Intuitively, it means that the strategy $\tt st$ transforms the hedge $\tt h1$ to the hedge $\tt h2$. (We will use this, somehow sloppy, but intuitively clear wording in this paper.) Its negation is written as \verb#st :: h1 =\=> h2#. A $\rho$Log \emph{literal} ($\rho$-literal) is a $\rho$-atom or its negation. A \prholog{} \emph{clause} is either a Prolog clause, or a clause of the form  \verb#st :: h1 ==> h2 :- body# (in the sequel called a $\rho$-clause) where \texttt{body} is a (possibly empty) conjunction of $\rho$- and Prolog literals.\footnote{In fact, \prholog\ clauses may have a more complex structure, when (some of) the literals are equipped with membership constraints, constraining possible values of sequence and context variables. Such constraints are taken into account in the matching process. For simplicity, we do not consider them in this paper.}

A \prholog\ \emph{program} is a sequence of \prholog\ clauses and a \emph{query} is a conjunction of $\rho$- and Prolog literals. A restriction on variable occurrence is imposed on clauses: If a $\rho$-clause has the body that contains Prolog literals, then the only variables that can occur in those Prolog literals are the \rholog\ individual variables. (When it comes to evaluating such Prolog literals, the individual variables there are converted into Prolog variables.) The same restriction applies to $\rho$-queries where Prolog literals occur. On the other hand, Prolog clauses can not contain any \rholog\ variables. In short: $\rho$-clauses and queries can contain only \rholog\ variables. Prolog clauses and queries can contain only Prolog variables.

Both a program clause and a query should satisfy a syntactic restriction, called well-modedness, to guarantee that each execution step is performed using matching (which is finitary in our language) and not unification (whose decidability is not known. It subsumes context unification whose decidability is a long-standing open problem~\cite{ContextUnif}.). To explain the essence of the problem, consider a query \verb#str :: (i_X, i_X) ==> i_X#. It contains (two copies of) a variable in the left-hand side, which might give rise to an arbitrarily complex context unification problem $t_1=^? t_2$, if there is a clause with the head of the form \verb#str :: (#$t_1$, $t_2$\verb#) ==> h# and $t_1$ and $t_2$ are terms containing context variables. It can be that the unification problem has infinitely many unifiers (this might be the case also with sequence variables), which leads to computing infinitely many answers. Even worse, since the decision algorithm for context unification is not known, an attempt to compute context unifiers might run forever without detecting that there are no more unifiers.

All these cases are extremely undesirable from the computational point of view. Therefore, we would like to restrict ourselves to the fragment that guarantees a terminating finitary solving procedure. Matching is one of such possible fragments. Well-moded clauses and queries forbid uninstantiated variables to appear in one of the sides of unification problems and, hence, allow only matching problems. Queries like \verb#str :: (i_X, i_X) ==> i_X# above and clauses that might lead to such kind of queries are not allowed in \prholog.

More specifically, well-modedness for \prholog\ programs extends the same notion for logic programs, introduced in~\cite{DembinskiMaluszinsky}: A mode for the relation ${\tt \cdot\,\verb#::#\,\cdot\, \verb#==>#\,\cdot}$ is a function that defines the input and output positions of the relation respectively as $\modein({\tt \cdot\,\verb#::#\,\cdot\, \verb#==>#\,\cdot})=\{1,2\}$ and $\modeout({\tt \cdot\,\verb#::#\,\cdot\, \verb#==>#\,\cdot})=\{3\}$. A mode is defined (uniquely) for a Prolog relation as well. A clause is moded if all its predicate symbols are moded. We assume that all $\rho$-clauses are moded. As for the Prolog clauses, we require modedness only for those ones that define a predicate that occurs in the body of some $\rho$-clause. If a Prolog literal occurs in a query in conjunction with a $\rho$-clause, then its relation and the clauses that define this relation are also assumed to be moded.

Before defining well-modedness, we introduce the notation $\vars(E)$ for a set of variables occurring in an expression $E$, and define $\vars(E,\{p_1,...,p_n\})=\cup_{i=1}^n \vars(E|_{p_i})$, where $E|_{p_i}$ is the standard notation for a subexpression of $E$ at position $p_i$. The symbol $\cVa$ stands for the set of anonymous variables. A ground expression contains no variables. Then well-modedness of queries and clauses are defined as follows:
\begin{defn} A query $\tt L_1,\ldots,L_n$ is well-moded iff it satisfies the following conditions for each $1\le i\le n$:
\begin{itemize}
  \item $\vars({\tt L_i}, \modein({\tt L_i}))\subseteq  \cup_{j=1}^{i-1}\vars({\tt L_j},\modeout({\tt L_j}))\setminus \cVa.$
  \item If $\tt L_i$ is a negative literal, then $\vars({\tt L_i}, \modeout({\tt L_i}))\subseteq  \cup_{j=1}^{i-1}\vars({\tt L_j},\modeout({\tt L_j}))\cup \cVa.$
  \item If $\tt L_i$ is a \rholog\ literal, then its strategy term is ground.
\end{itemize}
A clause $\tt L_0\pif L_1,\ldots,L_n$ is well-moded, iff the following conditions are satisfied for each $1\le i\le n$:
\begin{itemize}
  \item $\vars({\tt L_i}, \modein({\tt L_i}))\cup \vars({\tt L_0}, \modeout({\tt L_0}))\subseteq   \cup_{j=0}^{i-1}\vars({\tt L_j},\modeout({\tt L_j}))\setminus \cVa.$
  \item If $\tt L_i$ is a negative literal, then $\vars({\tt L_i}, \modeout({\tt L_i}))\subseteq \cup_{j=1}^{i-1}\vars({\tt L_j},\modeout({\tt L_j}))\cup \cVa \cup \vars({\tt L_0}, \modein({\tt L_0})).$
    \item If $\tt L_0$ and ${\tt L_i}$ are \rholog\ literals with the strategy terms $\tt st_0$ and $\tt st_i$, respectively, then $\vars({\tt st_i})\subseteq \vars({\tt st_0})$.
\end{itemize}
\end{defn}

\prholog\ allows only well-moded program clauses and queries. There is no restriction on the Prolog clauses if the predicate they define is not used in a $\rho$-clause.

\begin{exmp}\label{ex:well-moded}
 The query \verb#str1 :: a ==> i_X, str2 :: i_Y ==> i_Z# is not well-moded, because the variable \verb#i_Y# in the input position of the second subgoal does not occur in the output position of the first subgoal. On the other hand, \verb#str1 :: a ==> i_X, str2 :: i_X ==> i_Z# is well-moded.

 If we change the last goal into \verb#str1 :: a ==> i_X, str2 :: i_X =\=> i_Z#, well-modedness will get violated again, because the variable \verb#i_Z#, occurring in the negative literal, does not appear in the output position of the previous subgoal. Examples of well-moded queries involving negative literals are, e.g., \verb#str1 :: a ==> (i_X, i_Z),# \verb#str2 :: i_X =\=> i_Z # and \verb#str1 :: a ==> i_X,# \verb#str2 :: i_X =\=> i_#.
\end{exmp}

For well-moded programs and queries, \prholog\ uses Prolog's depth-first inference mechanism with the leftmost literal selection in the goal. If the selected literal is a Prolog literal, then it is evaluated in the standard way. If it is a $\rho$-atom of the form \verb#st :: h1 ==> h2#, then \prholog{} finds a (renamed copy of a) program clause \verb#st' :: h1' ==> h2' :-# \verb#body# such that $\tt st'$ matches $\tt st$ and $\tt h1'$ matches $\tt h1$ with a substitution $\sigma$. Then, it replaces the selected literal in the query with the conjunction of $\mathtt{body}\sigma$ and a literal that forces matching $\tt h2$ to $\tt h2'\sigma$, applies $\sigma$ to the rest of the query, and continues. Success and failure are defined in the standard way. Backtracking allows to explore other alternatives that may come from matching the selected query literal to the head of the same program clause in a different way, or to the head of another program clause.

Negative $\rho$-literals are processed by the standard negation-as-failure rule: A negative query of the form \verb#str :: h1 =\=> h2# succeeds if all attempts of satisfying its complementary literal, a positive query \verb#str :: h1 ==> h2#, end with failure. Well-modedness guarantees that whenever such a negative literal is selected during the \prholog{} execution process, there are no variables in it except, maybe, some anonymous variables that may occur in \verb#h2#.

\section{Strategic Programming in \prholog}
\label{sect:prog}

Strategies can be combined to express in a compact way many tedious small step transformations. These combinations give more control on transformations. \prholog{} provides a library of several predefined strategy combinators. Most of them are standard. The user can write her own strategies in \prholog{} or extend the Prolog code of the library. Some of the predefined strategies and their intuitive meanings are the following:
\begin{itemize}
  \item \verb#id :: h1 ==> h2# succeeds if the hedges \verb#h1# and \verb#h2# are identical (or can be made identical by \verb#h2# matching \verb#h1#) and fails otherwise.
  \item $\mathtt{compose(st_1,st_2,\ldots,st_n)}$, $n\ge 2$, first transforms the input hedge by $\mathtt{st_1}$ and then transforms the result by $\mathtt{compose(st_2,\ldots,st_n)}$ (or by $\mathtt{st_2}$, if $n=2$). Via backtracking, all possible results can be obtained. The strategy fails if either $\mathtt{st_1}$ or $\mathtt{compose(st_2,\ldots,st_n)}$ fails.
  \item $\mathtt{choice(st_1,\ldots,st_n)}$, $n\ge 1$, returns a result of a successful application of some strategy $\mathtt{st_i}$ to the input hedge. It fails if all $\mathtt{st_i}$'s fail. By backtracking it can return all outputs of the applications of each of the strategies $\mathtt{st_1,\ldots,st_n}$.
  \item \verb#first_one#$\mathtt{(st_1,\ldots,st_n)}$, $n\ge 1$, selects the first $\mathtt{st_i}$ that does not fail on the input hedge and returns only one result of its application. \verb#first_one# fails if all $\mathtt{st_i}$'s fail. Its variation, \verb#first_all#, returns via backtracking all the results of the application to the input hedge of the first strategy $\mathtt{st_i}$ that does not fail.
  \item $\mathtt{nf(st)}$, when terminates, computes a normal form of the input hedge with respect to \texttt{st}. It never fails because if an application of \texttt{st} to a hedge fails, then \texttt{nf(st)} returns that hedge itself. Backtracking returns all normal forms.
  \item $\mathtt{iterate(st, \mathit{N})}$ starts transforming the input hedge with \texttt{st} and returns a result (via backtracking all the results) obtained after $N$ iterations for a given natural number $N$.
  \item $\mathtt{map1(st)}$ maps the strategy \texttt{st} to each term in the input hedge and returns the result hedge. Backtracking generates all possible output hedges. \texttt{st} should operate on a single term and not on an arbitrary hedge. $\mathtt{map1(st)}$ fails if \texttt{st} fails on at least one term from the input hedge. $\mathtt{map}$ is a variation of $\mathtt{map1}$ where the single-term restriction is removed. It should be used with care because of high nondeterminism. Both $\mathtt{map1}$ and $\mathtt{map}$, when applied to the empty hedge, return the empty hedge.
  \item \texttt{interactive} takes a strategy from the user, transforms the input hedge by it and waits for further user instruction (either to apply another strategy to the result hedge or to finish).
   \item $\mathtt{rewrite(st)}$ applies to a single term (not to an arbitrary hedge) and rewrites it by \texttt{st} (which also applies to a single term). Via backtracking, it is possible to obtain all the rewrites. The input term is traversed in the leftmost-outermost manner. Note that $\mathtt{rewrite(st)}$ can be easily implemented inside \prholog{}: \\
       \verb#   rewrite(i_Str) :: c_Context(i_Redex) ==> c_Context(i_Contractum) :-#\\
       \verb#        i_Str :: i_Redex ==> i_Contractum.#
\end{itemize}

We give below few examples that demonstrate the use some of the \prholog{} features, including the strategies we just mentioned. The users can define own strategies in a program either by writing clauses for them or using abbreviations of the form \verb#str_1 := str_2#. Such an abbreviation stands for the clause \verb#str_1 :: s_X ==> s_Y :- str_2 :: s_X ==> s_Y#.
\begin{exmp}\label{ex:elementary}
  Let \verb#str1# and \verb#str2# be two strategies defined as follows:
  \begin{alltt}
    str1 :: (s_1, a, s_2) ==> (s_1, f(a), s_2).
    str2 :: (s_1, i_x, s_2, i_x, s_3) ==> (s_1, i_x, s_2, s_3).
  \end{alltt}\vspace{-0.4cm}

  Putting different strategies in the goal we get different answers:
  \begin{itemize}
    \item The goal \verb#str1 :: (a, b, a, f(a)) ==> s_X# returns two answers (instantiations of the sequence variable \verb#s_X#): \verb#(f(a), b, a, f(a))# and \verb#(a, b, f(a), f(a))#. Multiple answers are computed by backtracking. They are two because \verb#(s_1, a, s_2)# matches \verb#(a, b, a, f(a))# in two ways, with the matchers $\{$\verb#s_1#$\mapsto$\verb#eps#, \verb#s_2#$\mapsto$\verb#(b, a, f(a))#$\}$ and $\{$\verb#s_1#$\mapsto$\verb#(a, b)#, \verb#s_2#$\mapsto$\verb#f(a)#$\}$, respectively.
    \item If we change the previous goal into \verb#str1 :: (a, b, a, f(a)) ==> (s_X, f(a), s_Y)#, then \prholog{} will return four answers that correspond to the following instantiations of \verb#s_X# and \verb#s_Y#:
        \begin{enumerate}
          \item \verb#s_X#$\mapsto$\verb#eps#, \verb#s_Y#$\mapsto$\verb#(b, a, f(a))#.
          \item \verb#s_X#$\mapsto$\verb#(f(a), b, a)#, \verb#s_Y#$\mapsto$\verb#eps#.
          \item \verb#s_X#$\mapsto$\verb#(a, b)#, \verb#s_Y#$\mapsto$\verb#f(a)#.
          \item \verb#s_X#$\mapsto$\verb#(a, b, f(a))#, \verb#s_Y#$\mapsto$\verb#eps#.
        \end{enumerate}
    \item The goal \verb#str1 :: (a, b, a, f(a)) =\=> s_ # fails, because it's positive counterpart succeeds. On the other hand, \verb#str1 :: (a, b, a, f(a)) =\=> (b, s_)# succeeds.
    \item The composition \verb#compose(str1, str2) :: (a, b, a, f(a)) ==> s_X# gives two answers: \verb#(f(a), b, a)# and \verb#(a, b, f(a))#,
    \item On the goal \verb#choice(str1, str2) :: (a, b, a, f(a)) ==> s_X# we get three hedges as answers: \verb#(f(a), b, a, f(a))#, \verb#(a, b, f(a), f(a))#, and \verb#(a, b, f(a))#.
    \item \verb#nf(compose(str1, str2)) :: (a, b, a, f(a)) ==> s_X#, which computes a normal form of the composition, returns \verb#(f(a), b)# twice, computing it in two different ways.
    \item The goal \verb#first_one(str1, str2) :: (a, b, a, f(a)) ==> s_X# returns only one answer \verb#(f(a), b, a, f(a))#. This is the first output computed by the first applicable strategy, \verb#str1#.
    \item Finally, \verb#first_all(str1, str2) :: (a, b, a, f(a)) ==> s_X# computes two instantiations: \verb#(f(a), b, a, f(a))# and \verb#(a, b, f(a), f(a))#. These are all the answers returned by the first applicable strategy, \verb#str1#.
  \end{itemize}
\end{exmp}

\begin{exmp}\label{ex:flat}
  The two \prholog{} clauses below flatten nested occurrences of the head function symbol of a term. The code is written using function and sequence variables, which makes it reusable, since it can be used to flatten terms with different heads and different numbers of arguments:
  \begin{alltt}
    flatten_one :: f_Head(s_1, f_Head(s_2), s_3) ==> f_Head(s_1, s_2, s_3).
    flatten := nf(flatten_one).
  \end{alltt}\vspace{-0.4cm}

    The first clause flattens one occurrence of the nested head. The second one (written in the abbreviated form) defines the \verb#flatten# strategy as the normal form of \verb#flatten_one#. Here are some examples of queries involving these strategies:
    \begin{itemize}
    \item \verb#flatten_one :: f(a, f(b, f(c)), f(d)) ==> i_X# gives \verb#f(a, b, f(c), f(d))#.
    \item \verb#flatten :: f(a, f(b, f(c)), f(d)) ==> i_X# returns \verb#f(a, b, c, d)#.
    \item We can map the strategy \verb#flatten# to a hedge, which results in flattening each element of the hedge. For instance, the goal \verb#map1(flatten) :: (a, f(f(a)), g(a, g(b))) ==> s_X# returns the hedge \verb#(a, f(a), g(a, b))#.
    \end{itemize}
\end{exmp}

\begin{exmp}\label{ex:replace}
  The \verb#replace# strategy takes a term and a sequence of replacement rules, chooses a subterm in the given term that can be replaced by a rule, and returns the result of the replacement. \verb#replace_all# computes a normal form with respect to the given replacement rules.
  \begin{alltt}
    replace :: (c_Context(i_X), s_1, i_X -> i_Y, s_2) ==>
               (c_Context(i_Y), s_1, i_X -> i_Y, s_2).

    replace_all :: (i_Term, s_Rules) ==> i_Instance :-
                   nf(replace) :: (i_Term, s_Rules) ==> (i_Instance, s_).
  \end{alltt}\vspace{-0.4cm}

  With \verb#replace_all#, one can, for example, compute an instance of a term under an idempotent substitution: \verb#replace_all :: (f(x, g(x, y)), x -> z, y -> a) ==> i_X# gives \verb#f(z, g(z, a))#. (We can take the conjunction of this goal with the cut predicate to avoid recomputing the same instance several times.) The same code can be used to compute a normal form of a term under a ground rewrite system, the sort of a term if the rules are sorting rules, etc.
\end{exmp}

\begin{exmp}\label{ex:prove}
  This is a bit longer example that shows how one can specify a simple propositional proving procedure in \prholog{}. We assume that the propositional formulas are built over negation (denoted by `\verb#-#') and disjunction (denoted by `\verb#v#'). The corresponding \prholog{} program starts with the Prolog operator declaration that declares disjunction an infix operator:
  \begin{alltt}
   :- op(200, xfy, v).
  \end{alltt}\vspace{-0.4cm}

  Next, we describe inference rules of a Gentzen-like sequent calculus for propositional logic. The rules operate on sequents, represented as \verb#sequent(ant(#\emph{sequence of formulas}\verb#), cons(#\emph{sequence of formulas}\verb#))#. \verb#ant# and \verb#cons# are tags for the antecedent and consequent, respectively. There are five inference rules in the calculus: The axiom rule, negation left, negation right, disjunction left, and disjunction right.
  \begin{alltt}
   axiom :: sequent(ant(s_, i_Formula, s_), cons(s_, i_Formula, s_)) ==>  eps.

   neg_left :: sequent(ant(s_F1, -(i_Formula), s_F2), cons(s_F3)) ==>
      sequent(ant(s_F1, s_F2), cons(i_Formula, s_F3)).

   neg_right :: sequent(ant(s_F1), cons(s_F2, -(i_Formula), s_F3)) ==>
      sequent(ant(s_F1, i_Formula), cons(s_F2, s_F3)).

   disj_left :: sequent(ant(s_F1, i_Formula1 v i_Formula2, s_F2), i_Cons) ==>
      (sequent(ant(s_F1, i_Formula1, s_F2), i_Cons),
       sequent(ant(s_F1, i_Formula2, s_F2), i_Cons)).

   disj_right :: sequent(i_ant, cons(s_F1, i_Formula1 v i_Formula2, s_F2)) ==>
      sequent(i_ant, cons(s_F1, i_Formula1, i_Formula2, s_F2)).
  \end{alltt}\vspace{-0.3cm}

  Next, we need to impose control on the applications of the inference rules and define success and failure of the procedure. The control is pretty straightforward: To perform an inference step on a given hedge of sequents, we select the first sequent and apply to it the first applicable inference rule, in the order specified in the arguments of the strategy \verb#first_one# below. When there are no sequents left, the procedure ends with success. Otherwise, if no inference step can be made, we have failure. %
  \begin{alltt}
   success :: eps ==> true.

   inference_step :: (sequent(i_Ant, i_Cons), s_Sequents) ==>
                     (s_New_sequents, s_Sequents) :-
         first_one(axiom, neg_left, neg_right, disj_left, disj_right) ::
                     sequent(i_Ant, i_Cons) ==> s_New_sequents.

   failure :: (sequent(i_Ant, i_Cons), s_Sequents) ==> false.
  \end{alltt}\vspace{-0.4cm}

  Finally, we specify the proof procedure as repeatedly applying the first possible strategy between \verb#success#, \verb#inference_step#, and \verb#failure# (in this order) until none of them is applicable:
  \begin{alltt}
   prove := nf(first_one(success, inference_step, failure)).
  \end{alltt}\vspace{-0.4cm}

  Note that it does matter in which order we put the clauses for the inference rules or the control in the program. What matters, is the order they are combined (e.g. as it is done in the strategy \verb#first_one#).

  What we described here is just one way of implementing the given propositional proof procedure in \prholog{}. One could do it differently as well, for instance, by writing recursive clauses like it has been shown in \cite{MK06}. However, we believe that the version above is more declarative and naturally corresponds to the way the procedure is described in textbooks.
\end{exmp}

  Note that there can be several clauses for the same strategy in a \prholog{} program. In this case they behave as usual alternatives of each other (when a query with this strategy is being evaluated) and are tried in the order of their appearance in the program, top-down.

\section{Implementing Rewriting Strategies}
\label{sect:rewriting}

In this section we illustrate how rewriting strategies can be implemented in \prholog{}. It can be done in a pretty succinct and declarative way. The code for leftmost-outermost and outermost rewriting is shorter than the one for leftmost-innermost and innermost rewriting, because it takes an advantage of \prholog{}'s built-in term traversal strategy.

\paragraph{Leftmost-Outermost and Outermost Rewriting.}
\label{subsect:lo}
As mentioned above, the \texttt{rewrite} strategy traverses a term in leftmost-outermost order to rewrite subterms.
For instance, if the strategy \verb#strat# is defined by two rules
  \begin{alltt}
   strat :: f(i_X) ==> g(i_X).
   strat :: f(f(i_X)) ==> i_X.
  \end{alltt}\vspace{-0.3cm}
then for the goal \verb#rewrite(strat) :: h(f(f(a)), f(a)) ==> i_X# we get, via backtracking, four instantiations for \verb#i_X#, in this order: \verb#h(g(f(a)), f(a))#, \verb# h(a, f(a))#, \verb# h(f(g(a)), f(a))#, and \verb#h(f(f(a)), g(a))#.

If we want to obtain \emph{only one result}, then it is enough to add the cut predicate at the end of the goal: \verb#rewrite(strat) :: h(f(f(a)), f(a)) ==> i_X, !# returns only \verb#h(g(f(a)), f(a))#.

To get \emph{all the results of leftmost-outermost rewriting,} we have to find the first redex and rewrite it in all possible ways (via backtracking), ignoring all the other redexes. This can be achieved by using an anonymous variable for checking reducibility, and then putting the cut predicate:
  \begin{alltt}
   rewrite_left_out(i_Str) :: c_Context(i_Redex) ==> c_Context(i_Contractum) :-
        i_Str :: i_Redex ==> i_,
        !,
        i_Str :: i_Redex ==> i_Contractum.
  \end{alltt}

The goal \verb#rewrite_left_out(strat) :: h(f(f(a)), f(a)) ==> i_X# gives two instantiations for \verb#i_X#: \verb#h(g(f(a)), f(a))# and \verb#h(a, f(a))#.

To return \emph{all the results of outermost rewriting} we find an outermost redex and rewrite it. Backtracking returns all the results for all outermost redexes.
\begin{alltt}
   rewrite_out(i_Str) :: i_X ==> i_Y :-
        i_Str :: i_X ==> i_,
        !,
        i_Str :: i_X ==> i_Y.

   rewrite_out(i_Str) :: f_F(s_1, i_X, s_2) ==> f_F(s_1, i_Y, s_2) :-
        rewrite_out(i_Str) :: i_X ==> i_Y.
\end{alltt}

The goal \verb#rewrite_out(strat) :: h(f(f(a)), f(a)) ==> i_X# gives three answers, in this order: \verb#h(g(f(a)), f(a))#,\; \verb#h(a, f(a))#,\, and \verb#h(f(f(a)), g(a))#.

\paragraph{Leftmost-Innermost and Innermost Rewriting.}
\label{subsect:li}

Implementation of innermost strategy in \prholog{} is slightly more involved than the implementation of outermost rewriting. It is not surprising since the outermost strategy takes an advantage of the \prholog{} built-in term traversal strategy. For innermost rewriting, we could have modified the \prholog{} source by simply changing the order of two rules in the matching algorithm to give preference to the rule that descends deep in the term structure. It would change the term traversal strategy from leftmost-outermost to leftmost-innermost. Another way would be to build term traversal strategies into \prholog{} (like it is done in ELAN and Stratego, for instance) that would give the user more control on traversal strategies, giving her a possibility to specify the needed traversal inside a \prholog{} program.

However, here our aim is different: We would like to demonstrate that rewriting strategies can be implemented quite easily inside \prholog{}. For the outermost strategy it has already been shown. As for the innermost rewriting, if we want to obtain \emph{only one result by leftmost-innermost strategy}, we first check whether any argument of the selected subterm rewrites. If not, we try to rewrite the subterm and if we succeed, we cut the alternatives. The way how matching is done guarantees that the leftmost possible redex is taken:

\begin{alltt}
   rewrite_left_in_one(i_Str) :: c_Ctx(f_F(s_Args)) ==> c_Ctx(i_Contractum) :-
        rewrites_at_least_one(i_Str) :: s_Args =\verb|\|=> i_,
        i_Str :: f_F(s_Args) ==> i_Contractum,
        !.

   rewrites_at_least_one(i_Str) :: (s_, i_X, s_) ==> true :-
        rewrite(i_Str) :: i_X ==> i_,
        !.
\end{alltt}

To get \emph{all results of leftmost-innermost rewriting}, we first check whether the selected subterm is an innermost redex. If yes, the other redexes are cut off and the selected one is rewritten in all possible ways:

\begin{alltt}
   rewrite_left_in(i_Str) :: c_Context(f_F(s_Args)) ==>
                             c_Context(i_Contractum) :-
        rewrites_at_least_one(i_Str) :: s_Args =\verb|\|=> i_,
        i_Str :: f_F(s_Args) ==> i_,
        !,
        i_Str :: f_F(s_Args) ==> i_Contractum.
\end{alltt}

If \texttt{strat} is the strategy defined in the previous section, then we have only one answer for the goal \verb#rewrite_left_in(strat) :: h(f(f(a)), f(a)) ==> i_X#: the term \verb#h(f(g(a)), f(a))#. The same term is returned by \verb#rewrite_left_in_one#.

Finally, \verb#rewrite_in# computes \emph{all results of innermost rewriting} via backtracking:

\begin{alltt}
   rewrite_in(i_Str) :: f_F(s_Args) ==> i_Y :-
        rewrites_at_least_one(i_Str) :: s_Args =\verb|\|=> i_,
        i_Str :: f_F(s_Args) ==> i_Y.

   rewrite_in(i_Str) :: f_F(s_1, i_X, s_2) ==> f_F(s_1, i_Y, s_2) :-
        rewrite_in(i_Str) :: i_X ==> i_Y.
\end{alltt}

The goal \verb#rewrite_in(strat) :: h(f(f(a)), f(a)) ==> i_X# returns two instantiations of \verb#i_X#: \verb#h(f(g(a)), f(a))# and \verb#h(f(f(a)), g(a))#.

\section{Concluding Remarks}
\label{sect:concl}

\prholog{} extends Prolog with strategic conditional transformation rules that operate on hedges. The rules, written as clauses in \prholog{} programs, define strategies. Strategy combinators help the user to construct more complex strategies from simpler ones. \prholog{} queries may have several results. They can be explored by backtracking. Four different kinds of variables used in \prholog{} make the system expressive and flexible.

\prholog{} is based on Prolog's inference mechanism and allows Prolog clauses and predicates in its programs. The users familiar with logic programming and Prolog can very quickly start using \prholog{} since its syntax is similar to that of Prolog and semantics is based on logic programming.

We gave a brief overview on strategies in \prholog{}, explained some of them on examples, and showed how rewriting strategies can be compactly and declaratively implemented. \prholog{} is written in SWI-Prolog~\cite{swi} and has been tested for versions 5.6.50 and later. It is available for downloading from \verb#http://www.risc.uni-linz.ac.at/people/tkutsia/software.html#.

\section{Acknowledgments}
This research has been partially supported by the European Commission Framework 6 Programme for Integrated Infrastructures Initiatives under the project SCIEnce---Symbolic Computation Infrastructure for Europe (Contract No. 026133) and by JSPS Grant-in-Aid no. 20500025 for Scientific Research (C).

\bibliographystyle{eptcs}

\begin{thebibliography}{10}
\providecommand{\bibitemstart}[1]{\bibitem{#1}}
\providecommand{\bibitemend}{}
\providecommand{\bibliographystart}{}
\providecommand{\bibliographyend}{}
\providecommand{\url}[1]{\texttt{#1}}
\providecommand{\urlprefix}{Available at }
\providecommand{\bibinfo}[2]{#2}
\bibliographystart

\bibitemstart{ElanENTCS}
\bibinfo{author}{P.~Borovansky}, \bibinfo{author}{C.~Kirchner},
  \bibinfo{author}{H.~Kirchner}, \bibinfo{author}{P.~E. Moreau} \&
  \bibinfo{author}{M.~Vittek} (\bibinfo{year}{1996}):
  \emph{\bibinfo{title}{{ELAN:} A logical framework based on computational
  systems.}}
\newblock {\sl \bibinfo{journal}{Electronic Notes in Theoretical Computer
  Science}} \bibinfo{volume}{4}, pp. \bibinfo{pages}{35--50}.
\bibitemend

\bibitemstart{vandenBrand:2001}
\bibinfo{author}{M.~G.~J. van~den Brand}, \bibinfo{author}{A.~van Deursen},
  \bibinfo{author}{J.~Heering}, \bibinfo{author}{H.~A. de~Jong},
  \bibinfo{author}{M.~de~Jonge}, \bibinfo{author}{T.~Kuipers},
  \bibinfo{author}{P.~Klint}, \bibinfo{author}{L.~Moonen},
  \bibinfo{author}{P.~A. Olivier}, \bibinfo{author}{J.~Scheerder},
  \bibinfo{author}{J.~J. Vinju}, \bibinfo{author}{E.~Visser} \&
  \bibinfo{author}{J.~Visser} (\bibinfo{year}{2001}): \emph{\bibinfo{title}{The
  {{\sc Asf} + {\sc Sdf}} Meta-environment: {A}~Component-Based Language
  Development Environment}}.
\newblock In: \bibinfo{editor}{R.~Wilhelm}, editor: {\sl
  \bibinfo{booktitle}{Proceedings of the 10th International Conference on
  Compiler Construction ({CC'01})}}, {\sl \bibinfo{series}{LNCS}}
  \bibinfo{volume}{2027}. \bibinfo{publisher}{Springer}, pp.
  \bibinfo{pages}{365--370}.
\bibitemend

\bibitemstart{Claire}
\bibinfo{author}{Y.~Caseau}, \bibinfo{author}{F.-X. Josset} \&
  \bibinfo{author}{F.~Laburthe} (\bibinfo{year}{2002}):
  \emph{\bibinfo{title}{Claire: combining sets, search and rules to better
  express algorithms}}.
\newblock {\sl \bibinfo{journal}{Theory and Practice of Logic Programming}}
  \bibinfo{volume}{2}(\bibinfo{number}{6}), pp. \bibinfo{pages}{769–--805}.
\bibitemend

\bibitemstart{CirsteaKirchner01a}
\bibinfo{author}{H.~Cirstea} \& \bibinfo{author}{C.~Kirchner}
  (\bibinfo{year}{2001}): \emph{\bibinfo{title}{The rewriting calculus - {Part
  I}}}.
\newblock {\sl \bibinfo{journal}{Logic Journal of the IGPL}}
  \bibinfo{volume}{9}(\bibinfo{number}{3}), pp. \bibinfo{pages}{339--375}.
\bibitemend

\bibitemstart{CirsteaKirchner01b}
\bibinfo{author}{H.~Cirstea} \& \bibinfo{author}{C.~Kirchner}
  (\bibinfo{year}{2001}): \emph{\bibinfo{title}{The rewriting calculus - {Part
  II}}}.
\newblock {\sl \bibinfo{journal}{Logic Journal of the IGPL}}
  \bibinfo{volume}{9}(\bibinfo{number}{3}), pp. \bibinfo{pages}{377--410}.
\bibitemend

\bibitemstart{Maude}
\bibinfo{author}{M.~Clavel}, \bibinfo{author}{F.~Dur\'{a}n},
  \bibinfo{author}{S.~Eker}, \bibinfo{author}{P.~Lincoln},
  \bibinfo{author}{N.~Mart\'{\i}-Oliet}, \bibinfo{author}{J.~Meseguer} \&
  \bibinfo{author}{J.~F. Quesada} (\bibinfo{year}{2002}):
  \emph{\bibinfo{title}{Maude: specification and programming in rewriting
  logic}}.
\newblock {\sl \bibinfo{journal}{Theoretical Computer Science}}
  \bibinfo{volume}{285}(\bibinfo{number}{2}), pp. \bibinfo{pages}{187–--243}.
\bibitemend

\bibitemstart{DembinskiMaluszinsky}
\bibinfo{author}{P.~Dembinski} \& \bibinfo{author}{J.~Maluszynski}
  (\bibinfo{year}{1985}): \emph{\bibinfo{title}{{AND}-parallelism with
  intelligent backtracking for annotated logic programs}}.
\newblock In: {\sl \bibinfo{booktitle}{Proceedings of the 2nd {IEEE} Symposium
  on Logic Programming}}. \bibinfo{publisher}{IEEE Computer Society}, pp.
  \bibinfo{pages}{29--–38}.
\bibitemend

\bibitemstart{CHR}
\bibinfo{author}{T.~Fr\"{u}hwirth} (\bibinfo{year}{1998}):
  \emph{\bibinfo{title}{Theory and Practice of {Constraint Handling Rules}}}.
\newblock {\sl \bibinfo{journal}{J.~Logic Programming}}
  \bibinfo{volume}{37}(\bibinfo{number}{1--3}), pp. \bibinfo{pages}{95--138}.
\bibitemend

\bibitemstart{FutatsugiGJM85}
\bibinfo{author}{K.~Futatsugi}, \bibinfo{author}{J.~A. Goguen},
  \bibinfo{author}{J.-P. Jouannaud} \& \bibinfo{author}{J.~Meseguer}
  (\bibinfo{year}{1985}): \emph{\bibinfo{title}{Principles of {OBJ2}}}.
\newblock In: {\sl \bibinfo{booktitle}{Conference Record of the Twelfth Annual
  ACM Symposium on Principles of Programming Languages ({POPL'85})}}.
  \bibinfo{publisher}{ACM Press}, pp. \bibinfo{pages}{52--66}.
\bibitemend

\bibitemstart{FutatsugiN97}
\bibinfo{author}{K~Futatsugi} \& \bibinfo{author}{A.~T. Nakagawa}
  (\bibinfo{year}{1997}): \emph{\bibinfo{title}{An Overview of {CAFE}
  Specification Environment - An Algebraic Approach for Creating, Verifying,
  and Maintaining Formal Specifications over Networks}}.
\newblock In: {\sl \bibinfo{booktitle}{Proceedings of the First IEEE
  International Conference on Formal Engineering Methods {(ICFEM'97)}}}.
  \bibinfo{publisher}{IEEE Computer Society}, pp. \bibinfo{pages}{170--182}.
\bibitemend

\bibitemstart{Goguen00OBJ}
\bibinfo{author}{J.~A. Goguen}, \bibinfo{author}{T.~Winkler},
  \bibinfo{author}{K.~Futatsugi}, \bibinfo{author}{J.~Meseguer} \&
  \bibinfo{author}{J.-P. Jouannaud} (\bibinfo{year}{2000}):
  \emph{\bibinfo{title}{Introducing {OBJ}}}.
\newblock In: \bibinfo{editor}{J.~A. Goguen} \& \bibinfo{editor}{G.~Malcolm},
  editors: {\sl \bibinfo{booktitle}{Software Engineering with {OBJ} - Algebraic
  Specification in Action}}. \bibinfo{publisher}{Kluwer Academic Publishers},
  pp. \bibinfo{pages}{3--167}.
\bibitemend

\bibitemstart{KM05}
\bibinfo{author}{T.~Kutsia} \& \bibinfo{author}{M.~Marin}
  (\bibinfo{year}{2005}): \emph{\bibinfo{title}{Matching with Regular
  Constraints}}.
\newblock In: \bibinfo{editor}{G.~Sutcliffe} \& \bibinfo{editor}{A.~Voronkov},
  editors: {\sl \bibinfo{booktitle}{Logic in Programming, Artificial
  Intelligence and Reasoning. Proceedings of the 12th International Conference
  {LPAR'05}}}, {\sl \bibinfo{series}{LNAI}} \bibinfo{volume}{3835}.
  \bibinfo{publisher}{Springer}, pp. \bibinfo{pages}{215--229}.
\bibitemend

\bibitemstart{Lloyd87}
\bibinfo{author}{J.~Lloyd} (\bibinfo{year}{1987}):
  \emph{\bibinfo{title}{Foundations of Logic Programming}}.
\newblock \bibinfo{publisher}{Springer-Verlag}, \bibinfo{edition}{2nd} edition.
\bibitemend

\bibitemstart{MK06}
\bibinfo{author}{M.~Marin} \& \bibinfo{author}{T.~Kutsia}
  (\bibinfo{year}{2006}): \emph{\bibinfo{title}{Foundations of the Rule-Based
  System {$\rho$Log}}}.
\newblock {\sl \bibinfo{journal}{J.~Applied Non-Classical Logics}}
  \bibinfo{volume}{16}(\bibinfo{number}{1-2}), pp. \bibinfo{pages}{151--168}.
\bibitemend

\bibitemstart{MartiOlietMesseguer2001}
\bibinfo{author}{N.~Mart{\'\i}-Oliet} \& \bibinfo{author}{J.~Meseguer}
  (\bibinfo{year}{2002}): \emph{\bibinfo{title}{Rewriting Logic: Roadmap and
  Bibliography}}.
\newblock {\sl \bibinfo{journal}{Theoretical Computer Science}}
  \bibinfo{volume}{285}(\bibinfo{number}{2}), pp. \bibinfo{pages}{121--154}.
\bibitemend

\bibitemstart{ContextUnif}
\emph{\bibinfo{title}{{RTA List of Open Problems. Problem \#90. Are context
  unification and linear second order unification decidable?}}}
\newblock
  \bibinfo{howpublished}{http://rtaloop.mancoosi.univ-paris-diderot.fr/problem%
s/90.html}.
\bibitemend

\bibitemstart{swi}
\emph{\bibinfo{title}{{SWI}-Prolog}}.
\newblock \bibinfo{howpublished}{http://www.swi-prolog.org}.
\bibitemend

\bibitemstart{Visser01}
\bibinfo{author}{E.~Visser} (\bibinfo{year}{2001}):
  \emph{\bibinfo{title}{Stratego: {A} Language for Program Transformation Based
  on Rewriting Strategies}}.
\newblock In: \bibinfo{editor}{A.~Middeldorp}, editor: {\sl
  \bibinfo{booktitle}{Proceedings of the 12th International Conference on
  Rewriting Techniques and Applications ({RTA'01})}}, {\sl
  \bibinfo{series}{LNCS}} \bibinfo{volume}{256}. \bibinfo{publisher}{Springer},
  pp. \bibinfo{pages}{357--362}.
\bibitemend

\bibliographyend
\end{thebibliography}


\end{document}